\documentclass[aps,prd,superscriptaddress,showpacs,preprintnumbers,twocolumn]{revtex4}

\usepackage{graphicx}

\begin{document}

\preprint{BUHEP-07-01, CPT-P05-2007, DESY 07-008, WUB/07-01}

\title{Diquark correlations in baryons on the lattice with overlap quarks}

\author{Ronald Babich}
\affiliation{Department of Physics, Boston University, 590
Commonwealth Avenue, Boston MA 02215, USA}
\author{Nicolas Garron}
\affiliation{DESY, Platanenallee 6, D-15738 Zeuthen, Germany}
\author{Christian Hoelbling}
\affiliation{Department of Physics, Universit\"at\,Wuppertal,
Gaussstr.\,20, D-42119\,Wuppertal, Germany}
\author{Joseph Howard}
\affiliation{Department of Physics, Boston University, 590
Commonwealth Avenue, Boston MA 02215, USA}
\author{Laurent Lellouch}
\affiliation{Centre de Physique Th\'eorique,
CNRS Luminy, Case 907,
F-13288 Marseille Cedex 9,
France}
\thanks{CPT is ``UMR 6207 du CNRS et des universit\'es d'Aix-Marseille I,
d'Aix-Marseille II et du Sud Toulon-Var, affili\'ee \`a la FRUMAM.''}
\author{Claudio Rebbi}
\email[]{rebbi@bu.edu}
\affiliation{Department of Physics, Boston University, 590
Commonwealth Avenue, Boston MA 02215, USA}

\date{January 25, 2007}

\begin{abstract}
We evaluate baryon wave functions in both the Coulomb and Landau
gauge in lattice QCD.  These are constructed from quark propagators
calculated with the overlap Dirac operator on quenched gauge configurations at
$\beta=6$.  By comparing baryon states that differ in their diquark content,
we find evidence for enhanced correlation in the scalar diquark channel, as
favored by quark models.  We also summarize earlier results for diquark masses
in the Landau gauge, casting them in a form more easily compared with
subsequent studies.
\end{abstract}


\maketitle


\section{Introduction}

The notion of a diquark is nearly as old as that of quarks themselves
and has been invoked to explain many aspects of hadron phenomenology
(see~\cite{Anselmino:1992vg} for a review).  Most generally, a diquark
is any two quark system, but the term is more often taken to denote two
correlated quarks in a particular representation of flavor and spin.  In
QCD-inspired quark models~\cite{DeRujula:1975ge}, the color-hyperfine
interaction gives rise to attraction in the spin singlet, SU(3)-flavor
anti-triplet
channel, a configuration known as a scalar diquark or more evocatively as a
``good'' diquark.  In contrast, the spin triplet, flavor sextet channel is
repulsive, and the associated axial vector or ``bad'' diquark
is disfavored.  Note that in this discussion and the rest of the paper, we
only consider positive-parity diquarks in the $\bar{3}$ of color, as would
describe two valence quarks in a baryon.  While one may write down diquark
operators symmetric in color, all evidence points toward their being
energetically disfavored.

In recent years, diquarks have received increased attention in light of the
possible existence of exotic states such as the $\Theta^+$, as diquark models
make definite predictions for their properties~\cite{Jaffe:2003sg}.  The
status of the $\Theta^+$ remains uncertain (see~\cite{Battaglieri:2006vu}
for a recent review of the experimental situation), but it serves
to remind one of the relative lack of other exotics naively allowed by QCD,
a scarcity that may largely be explained if diquark correlations play an
important role in hadron structure~\cite{Jaffe:2004ph}.

Ideally, issues such as these should be addressed by direct appeal to the
fundamental theory.  The lattice is the principal calculational framework
for nonperturbative QCD and has been brought to bear on the question of
diquarks in several recent studies (we set aside direct searches for exotic
states).  
Perhaps the most straightforward approach is to construct a diquark two-point
function and consider its fall-off in time, as one does to extract hadron
masses.  A diquark by itself is not a color singlet, however, and so
one must either fix the gauge or introduce an additional source of color.
The former approach was first pursued in~\cite{Hess:1998sd}, where diquark
correlators were calculated with Wilson fermions in the Landau gauge.  More
recently, we presented a similar investigation in~\cite{Babich:2005ay}
with overlap fermions at significantly lighter quark masses.  By comparing the
effective mass of the diquark with that of its constituent quarks, the scalar
diquark was found to be bound in the limit of vanishing quark mass.
In Section~\ref{masses} below, we briefly summarize these results in order
to give values for mass splittings that may be more easily compared with
subsequent studies.  In the second approach, one constructs a gauge invariant
object by contracting the free color index of the diquark at source and sink
with a Wilson line, serving as a static
quark~\cite{Orginos:2005vr,Alexandrou:2006cq}.  This allows one to extract
diquark mass differences, in qualitative agreement with the fixed-gauge
approach.  See also~\cite{Liu:2006zi}, where point-to-point baryon correlators
containing various diquarks are compared to those in the free theory.

While useful, such mass determinations provide limited information
about the nature of diquark correlations.  In this work, we directly
investigate spatial correlations among quarks in baryons by calculating
baryon wave functions on the lattice.  At least two natural formalisms exist
for defining what is meant by a ``wave function.''  The one pursued here
begins with a standard baryon correlator and involves displacing quarks at the
sink.  This function of quark displacements is then evaluated in a fixed
gauge~\cite{Velikson:1984qw}.  A very early study
of such wave functions may be found in~\cite{Gottlieb:1985xq} and more
complete investigations in~\cite{Hecht:1992uq,Hecht:1992ps}.  These
treat only a subset of all possible quark displacements and are largely
motivated by a desire for improved interpolating operators for spectroscopy.
Nevertheless, and although not emphasized, the nucleon wave function
parametrized in~\cite{Hecht:1992uq,Hecht:1992ps} does exhibit characteristics
attributable to diquark effects, in particular a negative charge radius
for the neutron.

An alternative definition of a hadronic wave function is that provided
by the density-density correlator
method~\cite{Barad:1984px,Barad:1985qd,Wilcox:1986dk}.  A recent addition
to the body of work treating such correlation functions for
baryons~\cite{Chu:1990ps,Chu:1994vi,Burkardt:1994pw} may be found
in~\cite{Alexandrou:2002nn}, where the focus is on possible deformations
arising from spin-orbit coupling.  Finally, a very recent
study~\cite{Alexandrou:2006cq} employs the density-density correlator
technique to examine the wave function of a diquark constrained to a
spherical shell about a static quark, a gauge-invariant setup mentioned
above in the context of diquark mass differences.  By fitting to an
exponential ansatz, the authors of~\cite{Alexandrou:2006cq} find a large,
but finite, radius for the scalar diquark in this environment.

In this work, we present the first detailed study of diquark correlations in
physical baryons (with all quark masses finite).  We consider all possible
displacements of the three quarks and calculate wave functions in both the
Coulomb and Landau gauges.  By directly comparing wave functions of disparate
states and calculating ratios of mean quark separations, we find evidence
of enhanced correlation in the scalar diquark channel.  We work in quenched
QCD and employ the overlap Dirac
operator~\cite{Neuberger:1998bg,Neuberger:1998fp,Narayanan:1995gw,
Narayanan:1994sk} in our calculation, a discretization which preserves
chiral symmetry on the lattice~\cite{Ginsparg:1982bj,Luscher:1998pq} and is
thereby closest to the continuum formulation.

The paper is organized as follows.  In Section~\ref{method}, we provide
details of our calculation and describe the correlation functions and
states that we study.  In Section~\ref{bwf} and its subsections, we
present and compare our baryon wave functions and from them calculate values
for mean quark separations.  Finally, in Section~\ref{masses}, we calculate
diquark mass differences from data first presented in~\cite{Babich:2005ay}.


\section{\label{method}Details of the calculation}

This study is one in a series employing the overlap Dirac operator on
a large lattice.  Results for meson and baryon spectra, as well as meson wave
functions, diquark correlators, and other observables were presented
in~\cite{Babich:2005ay}.  In~\cite{Babich:2006bh}, we calculated matrix
elements relevant for kaon physics in the standard model and beyond with a
careful treatment of nonperturbative renormalization in the RI/MOM scheme.
We direct the reader to~\cite{Babich:2005ay} for a discussion of the many
advantages of the overlap discretization as well as for details of our
implementation beyond those given here.

The overlap Dirac operator describing a massless quark~\cite{Neuberger:1998fp}
is given by
\begin{equation}
D=\frac{\rho}{a}\left(1+\frac{X}{\sqrt{X^\dagger X}}\right)\,,
\end{equation}
where $X=D_W-\rho/a$ is the Wilson Dirac operator with mass $-\rho/a$.
It follows that inversion of the overlap operator requires the
repeated calculation of $1/\sqrt{X^\dagger X}$.  This is accomplished with
polynomial or rational function approximations and is very demanding
computationally.  An unquenched calculation on a lattice as large as ours
would be beyond the capability of presently available resources.  We
therefore work in the quenched approximation and note that prior experience
with Wilson fermions has shown hadronic wave functions of the type we study
to be largely unaffected by quenching~\cite{Hecht:1992ps}.

We employ the Wilson gauge action with $\beta=6$ on a lattice of size
$18^3\times 64$.  This gives an inverse lattice spacing $a^{-1}$ of
2.12~GeV~\cite{Guagnelli:1998ud,Necco:2001xg} on the basis of the Sommer
scale defined by $r_0^2 F(r_0)=1.65$ with $r_0=0.5$~fm~\cite{Sommer:1993ce}.
One hundred independent gauge configurations were generated and then fixed
to the Landau gauge before inverting the Dirac operator.  The negative
mass parameter in the definition of the overlap was set to $\rho=1.4$ in
order to maximize locality~\cite{Hernandez:1998et}.  Quark propagators were
calculated from a point source for all color-spin combinations with a
conjugate gradient multimass solver for bare quark masses
$am_q=0.03,0.04,0.06,0.08,0.10,0.25,0.50,0.75$.  For reference, corresponding
values of the pion mass as well as baryon masses are given in
Table~\ref{tab-masses}~\cite{Babich:2005ay}.

\begin{table}[b]
\begin{tabular}{@{\extracolsep{1em}}llll}
\hline\hline
$am_q$ & $aM_P$ & $aM_8$ & $aM_{10}$ \\
\hline
0.03 & 0.219(3) & 0.63(2) & 0.75(3) \\
0.04 & 0.247(2) & 0.66(2) & 0.78(2) \\
0.06 & 0.297(2) & 0.714(11) & 0.82(2) \\
0.08 & 0.340(2) & 0.763(9) & 0.868(12) \\
0.10 & 0.3803(14) & 0.810(7) & 0.909(10) \\
\hline\hline
\end{tabular}
\caption{\label{tab-masses}Masses, in lattice units, of the lightest
pseudoscalar meson, octet baryon (e.g.~nucleon), and decuplet baryon for
quarks of equal mass $am_q$.  Quoted errors are statistical only.}
\end{table}

We now describe the states that we study. We work in a Dirac basis where
$\gamma_4$ is diagonal and utilize ``nonrelativistic'' wave functions
involving only either upper or lower spinor components.  Labeling the three
quarks $u,d,s$ for convenience, we give the spin structure of the states
of interest in Table~\ref{tab-states} in a transparent notation.  In the SU(3)
classification, these correspond to the octet $\Lambda$ and $\Sigma$ states
and the decuplet $\Sigma^*$.  In Section~\ref{bwf}, we will find it most
illuminating to compare the $\Lambda$ to the $\Sigma^*$; in the former, the
$u$ and $d$ are in a spin-0 ``good diquark'' configuration, while in the
latter they are in a spin-1.  The octet $\Sigma$ is a cousin of the nucleon
in which the pairs $u,s$ and $d,s$ are in superpositions of spin-0 and spin-1.

\begin{table}[t]
\begin{tabular}{l@{\hspace{1em}}|@{\hspace{1em}}c}
\hline\hline
$\Lambda$ & $(u_\uparrow d_\downarrow s_\uparrow 
            - u_\downarrow d_\uparrow s_\uparrow)/\sqrt{2}$ \\
          & $(u_\downarrow d_\uparrow s_\downarrow 
            - u_\uparrow d_\downarrow s_\downarrow)/\sqrt{2}$ \\
\hline
$\Sigma$ & $(u_\uparrow d_\downarrow s_\uparrow
             + u_\downarrow d_\uparrow s_\uparrow
            -2 u_\uparrow d_\uparrow s_\downarrow)/\sqrt{6}$ \\
         & $(u_\downarrow d_\uparrow s_\downarrow
             + u_\uparrow d_\downarrow s_\downarrow
            -2 u_\downarrow d_\downarrow s_\uparrow)/\sqrt{6}$ \\
\hline
$\Sigma^*$ & $u_\uparrow d_\uparrow s_\uparrow$ \\
           & $(u_\downarrow d_\uparrow s_\uparrow
             + u_\uparrow d_\downarrow s_\uparrow
             + u_\uparrow d_\uparrow s_\downarrow)/\sqrt{3}$ \\
           & $(u_\uparrow d_\downarrow s_\downarrow
             + u_\downarrow d_\uparrow s_\downarrow
             + u_\downarrow d_\downarrow s_\uparrow)/\sqrt{3}$ \\
           & $u_\downarrow d_\downarrow s_\downarrow$ \\
\hline\hline
\end{tabular}
\caption{\label{tab-states}Baryon states.}
\end{table}

For a given state, with spin structure as given in the table, we construct
a zero-momentum correlator
\begin{eqnarray}
G(\vec{r}_u,\vec{r}_d,t)&=&\sum_{\vec{r}} \langle u(\vec{r}+\vec{r}_u,t)
d(\vec{r}+\vec{r}_d,t) s(\vec{r},t) \nonumber\\
&&{}\times \bar u(\vec{0},0) \bar d(\vec{0},0) \bar s(\vec{0},0) \rangle \,.
\label{eq-corr}
\end{eqnarray}
Here color indices are implicit and are contracted with the antisymmetric
tensor at source and sink.  We combine correlators for the (two for the octet,
four for the decuplet) spin states distinguished by $J_z$.  Finally, to the
forward-propagating correlators constructed with upper spinor components
we add correlators propagating in the backward time direction that have been
constructed with lower components.  We thereby double our statistics while
ensuring that only the desired positive-parity states are excited from the
vacuum.

\begin{figure}[b]
\includegraphics*[width=4cm]{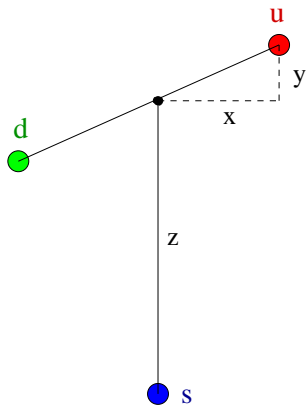}
\caption{\label{fig-geom}Geometry for visualizing the wave functions.}
\end{figure}

Since the quarks at the sink may be taken to be at distinct spatial sites,
Eq.~(\ref{eq-corr}) is only well-defined if we specify the gauge.  In
Section~\ref{bwf} we evaluate this correlation function in both the Coulomb
and Landau gauges.  Coulomb gauge-fixing was performed using simulated
annealing, starting from gauge configurations already fixed to the Landau
gauge.  At sufficiently large times, $G(\vec{r}_u,\vec{r}_d,t)$ settles into
a spatial profile that is independent of $t$ up to normalization.  We refer
to this profile as the ``wave function,''
\begin{equation}
\Psi(\vec{r}_u,\vec{r}_d) = \frac{G(\vec{r}_u,\vec{r}_d,t)}
{\sqrt{\sum_{\vec{r}_u,\vec{r}_d} |G(\vec{r}_u,\vec{r}_d,t)|^2}}\,.
\end{equation}
This zero-momentum wave function in general depends on two 3-vectors,
i.e.~six numbers.  As discussed in the next section, however, we only resolve
a dependence on separations between pairs of quarks, and it is therefore
effectively a function of a triangle, parametrized by three numbers.
For the purpose of displaying the wave function, we adopt the geometry
shown in Fig.~\ref{fig-geom}.  Here $z$ is the distance between the
quark labeled by $s$ and the center of mass of $u$ and $d$.  The axis
determined by $z$ is taken to establish a coordinate system in which we
specify the position $(x,y)$ of $u$ with respect to the center of mass.
We note that the states we consider are all symmetric under interchange
of the positions of $u$ and $d$.  In summary, the wave function in these
coordinates is given by
\begin{eqnarray}
\Psi(x,y,z)&=&\sum_{\vec{r}_u}\sum_{\vec{r}_d} \frac{\Psi(\vec{r}_u,\vec{r}_d)}
{\Psi(\vec{0},\vec{0})}
\delta\left(z-\frac{1}{2}|\vec{r}_u+\vec{r}_d|\right) \nonumber \\
&&{}\times\delta\left(y-\frac{(\vec{r}_u-\vec{r}_d)\cdot (\vec{r}_u+\vec{r}_d)}
   {2 |\vec{r}_u+\vec{r}_d|}\right) \nonumber \\
&&{}\times\delta\left(x-\sqrt{\frac{1}{4}|\vec{r}_u-\vec{r}_d|^2 - y^2}\right)\,,
\end{eqnarray}
where we have normalized the amplitude to unity where all three quarks are at
the same site and have defined the delta function on the lattice taking into
account the multiplicity of the sites.  In constructing the wave function, we
only consider configurations of the quarks where no two are separated by more
than half the length of the lattice ($L/2=9a$).

We conclude this section with some final details of our implementation.
First, we note that it might be advantageous to replace the point source
in Eq.~(\ref{eq-corr}) with an extended operator that better overlaps
the desired state.  We were constrained in our calculation, however, by the
fact that point-source propagators were required for studies of
nonperturbative renormalization and weak matrix elements; the calculation
of an additional set of smeared-source propagators was deemed too costly to be
worthwhile.

The wave functions we present in Section~\ref{bwf} were calculated by summing
over all possible positions of the three quarks.  Since each sum is over
$18^3$ sites, this involves a nontrivial amount of work.  We were able to
greatly speed up the calculation, however, by employing a fast Fourier
transform and utilizing the convolution theorem to eliminate one of the
summations.  A related issue is the large amount of data that would have to be
stored to capture all possible relative displacements of the quarks (i.e.~all
possible embeddings of a triangle in the lattice).  This was avoided by
adopting the parametrization described above and building a histogram in the
$x,y,z$ coordinates with linear interpolation.  The bin size was taken
to be $0.225a$ in $x,y$ and $0.45a$ in $z$, sufficiently small that the mean
quark separations presented in Section~\ref{sep} are unbiased, as confirmed
by examining the totally symmetric $\Sigma^*$ state.

Computations were performed with shared memory code on IBM p690 systems at
Boston University and NCSA.


\section{\label{bwf}Baryon wave functions}


\subsection{Wave functions}

Visualization of the wave functions will prove to be quite useful
for discerning differences in spatial correlations between states.  As a first
step, we must choose the time $t$ at which to evaluate the wave function.
For small times, the correlator $G(\vec{r}_u,\vec{r}_d,t)$ in
Eq.~(\ref{eq-corr}) is dominated by excited states.  It is therefore
necessary to take $t$ sufficiently large that the spatial profile has
settled into that of the ground state.  We find that for the states we study,
the wave function has settled by $t=8a$, in agreement with what was observed
for effective masses when calculating baryon spectra~\cite{Babich:2005ay}.
We conservatively take $t=10a$ in the remainder of this paper.

\begin{figure}
\includegraphics*[width=8.5cm]{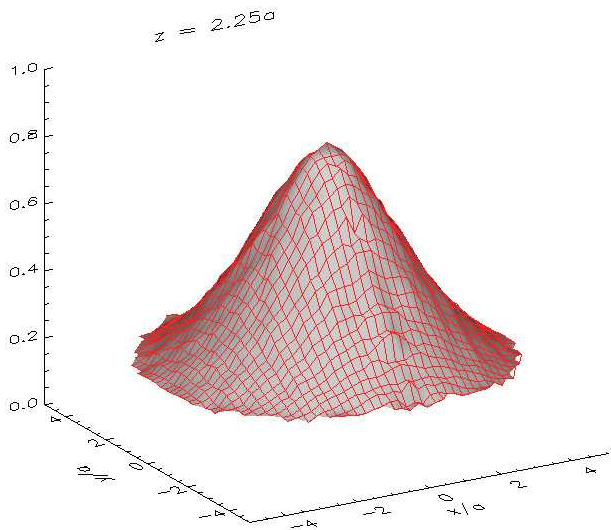}
\caption{\label{fig-octa-z5-landau}(Color online) Wave function of the
$\Lambda$ evaluated at $t=10a$ in the Landau gauge, for $z=2.25a$.}
\end{figure}

\begin{figure}
\includegraphics*[width=8.5cm]{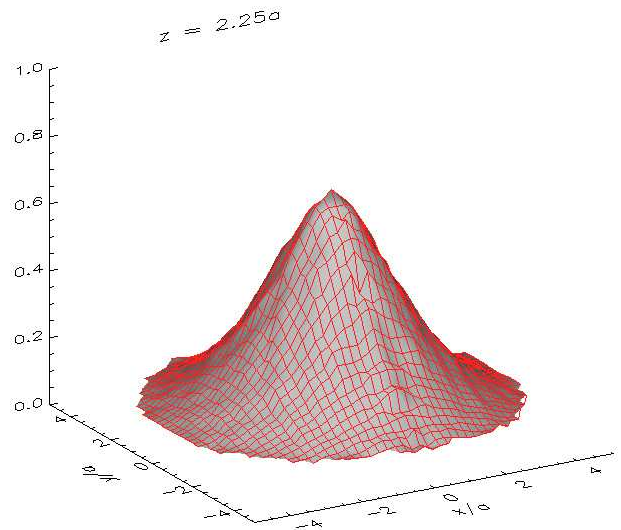}
\caption{\label{fig-octa-z5-coulomb}(Color online) Wave function of the
$\Lambda$ evaluated at $t=10a$ in the Coulomb gauge, for $z=2.25a$.}
\end{figure}

For plotting purposes, we fix $z$, the distance between the center of mass
of the first two quarks and the position of the third.  In
Fig.~\ref{fig-octa-z5-landau}, we plot the $\Lambda$ wave function in the
Landau gauge as a function of $x,y$ for one such $z$ separation.
All three quark masses are taken to be $am_q=0.03$, the lightest available
value.  The corresponding wave function in the Coulomb gauge is plotted in
Fig.~\ref{fig-octa-z5-coulomb}.  Recall that the wave functions have been
normalized to 1 where all three quarks are at the same site ($x=y=z=0$, not
shown).  Figures~\ref{fig-octa-z5-landau} and~\ref{fig-octa-z5-coulomb}
exhibit the general property that Coulomb-gauge wave functions are less broad
and better contained in the lattice volume than those calculated in the
Landau gauge, in agreement with~\cite{Hecht:1992uq,Hecht:1992ps}.
We will focus on Coulomb-gauge wave functions in the remainder of this section.

Statistical errors, which for clarity are not shown in
Figs.~\ref{fig-octa-z5-landau}-\ref{fig-both-z10}, are on the order of 6 to 10
percent (see Fig.~\ref{fig-xsec} below).  It is noteworthy that the overall
amplitude of the wave function tends to vary configuration by configuration
while it maintains the same basic shape.  In other words, if the wave function
at its peak (always at $\vec{r}_u=\vec{r}_d=0$) is found to be larger than
average on a given gauge configuration, it is likely to be larger at all
other quark displacements on that configuration.
In~\cite{Hecht:1992uq,Hecht:1992ps}, this effect was taken
as motivation to normalize the wave functions on a per-configuration basis.
While effective, this approach is difficult to justify from a field-theoretic
perspective and we do not pursue it here.  We note, however, that when
comparing the properties of various wave functions quantitatively, such
contributions to the errors often cancel, as we find for mean quark
separations in the next section.

By parametrizing our wave functions in terms of relative separations,
without regard to orientation, we have implicitly assumed isotropy.
Of course, one recognizes that there is a preferred direction, the
$z$-direction of the lattice (not to be confused with our $z$ coordinate)
with respect to which the $z$-component of spin is defined.  To test for the
possible presence of spin-orbit coupling, we added a fourth dimension to
our histogram with the new variable being the projection of the vector
whose length we call ``$z$'' along the $z$-direction of the lattice.  We
then constructed a wave function with definite $J_z$ and looked for a
dependence on this variable.  Within errors, we found no evidence for such a
dependence.  We conclude that the effects of spin-orbit coupling, if present,
are below the statistical limits of our calculation.

We come now to the main point of interest.  For the  $\Lambda$ state whose
wave function is plotted in Fig.~\ref{fig-octa-z5-coulomb}, the $u,d$
quarks are in the spin-0, ``good diquark'' configuration.  We would like to
compare this wave function to that of the $\Sigma^*$, where the two quarks
are in the spin-1 configuration.  In Fig.~\ref{fig-both-z5} and
Fig.~\ref{fig-both-z10}, we plot the two wave functions together for
two different $z$ separations.  A cross-section of Fig.~\ref{fig-both-z5} with
$y=0$ is shown in Fig.~\ref{fig-xsec} to give an indication of the errors.
As predicted in the literature, we note significantly stronger clustering
when $u,d$ are in the good diquark configuration.  This feature is independent
of $z$.

\begin{figure}
\includegraphics*[width=8.5cm]{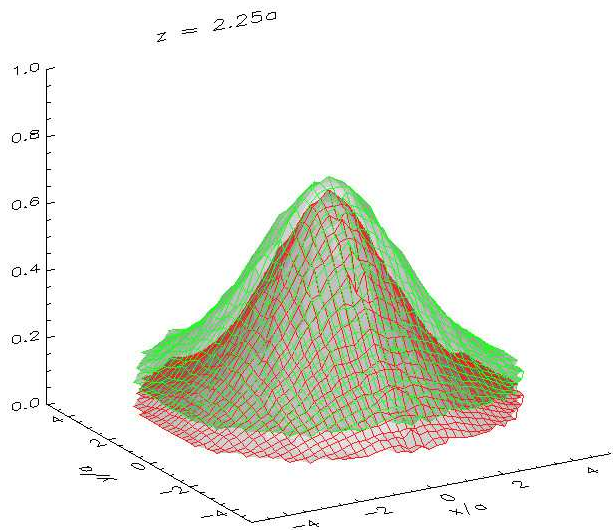}
\caption{\label{fig-both-z5}(Color online) Comparison of $\Lambda$ (red) and
$\Sigma^*$ (broader, in green) wave functions in the Coulomb gauge,
for $z=2.25a$.}
\end{figure}

\begin{figure}
\includegraphics*[width=8.5cm]{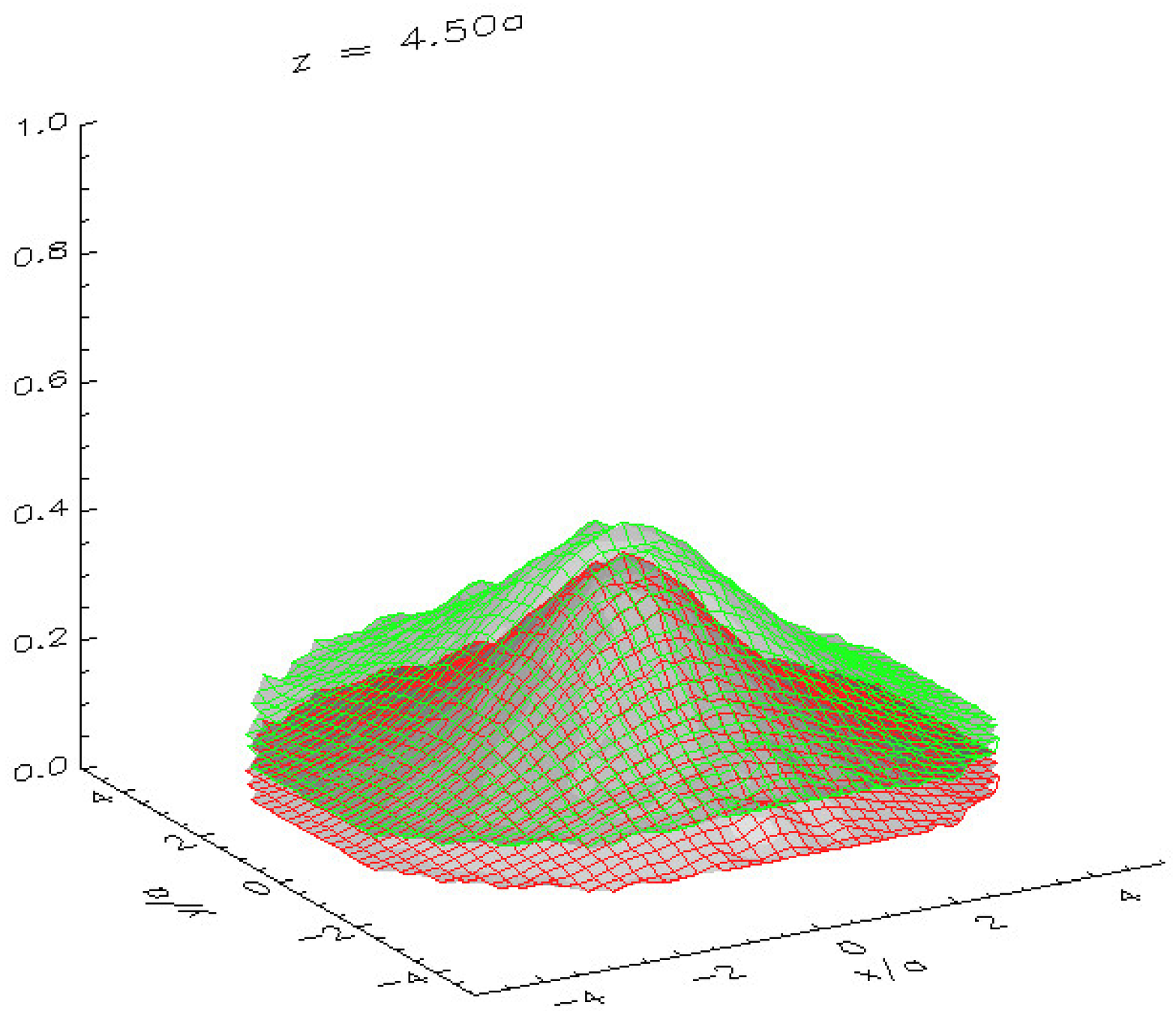}
\caption{\label{fig-both-z10}(Color online) Comparison of $\Lambda$ (red) and
$\Sigma^*$ (broader, in green) wave functions in the Coulomb gauge,
for $z=4.50a$.}
\end{figure}

\begin{figure}
\includegraphics*[width=8cm,angle=270]{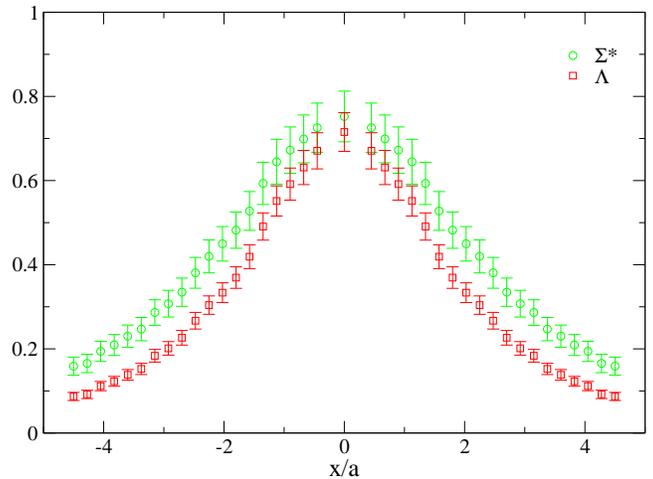}
\caption{\label{fig-xsec}(Color online) Cross-sections of $\Lambda$ and
$\Sigma^*$ wave functions with $z=2.25a$ and $y=0$.}
\end{figure}

In this section, we have presented results for baryons where the three quarks
are taken to be degenerate in mass with $am_q=0.03$.  In the next section,
we will evaluate the effect of increasing this mass.  One may also consider
baryons where the $s$ quark is taken to be significantly heavier than
the others.  None of the qualitative features are changed, but we do observe
a slight tendency for the mean separation between the two light quarks to be
larger than that between one of the light quarks and the heavy quark, when
considering the otherwise symmetric $\Sigma^*$ state.  This is a purely
kinematic effect that would apply even in a classical system of one heavy
and two light particles bound by two-body interactions.


\subsection{\label{sep}Mean quark separations}

From our wave functions, we calculate the mean square separation between
the $u$ and $d$ quarks in the natural way:
\begin{equation}
\langle |\vec{r}_u-\vec{r}_d|^2\rangle
=\sum_{\vec{r}_u}\sum_{\vec{r}_d} |\Psi(\vec{r}_u,\vec{r}_d)|^2\,
|\vec{r}_u-\vec{r}_d|^2 \,.
\end{equation}
Similarly, noting that our coordinates are defined such that $\vec{r}_s=0$,
\begin{equation}
\langle |\vec{r}_u-\vec{r}_s|^2\rangle
=\sum_{\vec{r}_u}\sum_{\vec{r}_d} |\Psi(\vec{r}_u,\vec{r}_d)|^2\,|\vec{r}_u|^2
\,.
\end{equation}
Since our baryons reside in a finite volume, such mean separations
must be interpreted with care.  In particular, we do not take into
account the tails of the wave functions that extend into adjacent cells
of our periodic lattice, nor do we remove those that impinge from them.
To do so would require that we model and fit the numerical wave
functions.  In contrast, the separations that we calculate follow
directly from the data. In the large volume limit, these separations
would converge to definite values.  In our finite volume, they provide a
rough quantitative estimate of the clustering observed in the scalar
diquark channel and of the dependence of such clustering on quark mass.
To the extent that finite volume effects are present, they are expected
only to weaken correlations.
In Tables~\ref{tab-sep-landau} and~\ref{tab-sep-coulomb} in the appendix,
we collect root mean square separations for the various states in the
two gauges. Errors have been calculated via the bootstrap method with 500
samples.

In QCD-inspired quark models~\cite{DeRujula:1975ge}, the Hamiltonian
generally includes a term of the form 
\begin{equation}
H=\alpha_s c \sum_{i<j} \frac{1}{m_i m_j} \vec{s}_i\cdot\vec{s}_j
\end{equation}
that is attractive in the spin singlet channel.  Here $\vec{s}_i$ is the
spin and $m_i$ the (constituent) mass of the $i$th quark, and $c$ is a
constant.  It follows that the strength of the interaction increases as quark
masses decrease.  For the purpose of quantifying the mass dependence of our
wave functions, we define a ratio of RMS separations,
\begin{equation}
{\cal R}^{ud}_{us}=\sqrt{\frac{\langle |\vec{r}_u-\vec{r}_d|^2\rangle}
   {\langle |\vec{r}_u-\vec{r}_s|^2\rangle}} \,.
\end{equation}
As noted earlier, fixed-gauge wave functions are generally broader in the
Landau gauge than in the Coulomb gauge.  For example, for the $\Lambda$ state
with $am_q=0.03$ in the Landau gauge, we find
$\sqrt{\langle |\vec{r}_u-\vec{r}_d|^2\rangle}=5.63(6)$, as compared to
5.17(9) in the Coulomb gauge.  Remarkably, however, ratios of separations
appear to be rather independent of gauge.

\begin{figure}
\includegraphics*[width=8.5cm]{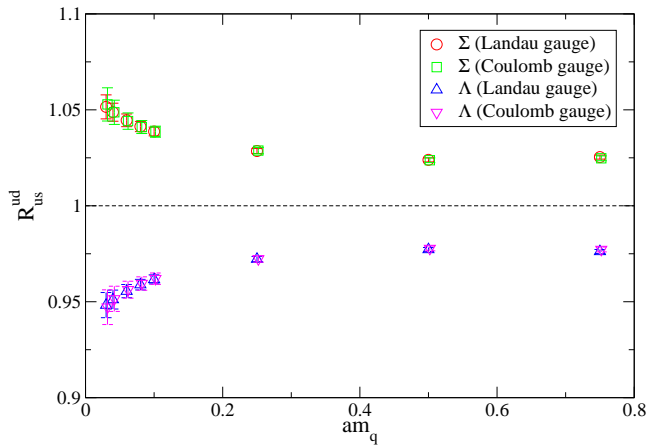}
\caption{\label{fig-R}(Color online) Ratios of RMS quark separations as a
function of bare quark mass $am_q$.}
\end{figure}

In Fig.~\ref{fig-R}, we plot such ratios in both gauges for the two octet
states and all available quark masses.  By construction, the decuplet
$\Sigma^*$ is totally symmetric; the corresponding ratio is exactly one and
would lie on the dotted line in the figure.  We recall that in the $\Lambda$,
the $u,d$ are in the spin-0 configuration while in the $\Sigma$, the
$u,s$ and $u,d$ are in superpositions of spin-0 and spin-1.  Again,
errors have been calculated with the bootstrap and leave little doubt that
spatial correlations are enhanced in the scalar diquark channel.  We also
observe that the effect strengthens markedly at the lightest masses.

\section{\label{masses}Diquark masses}

\begin{figure}
\includegraphics*[width=8.5cm]{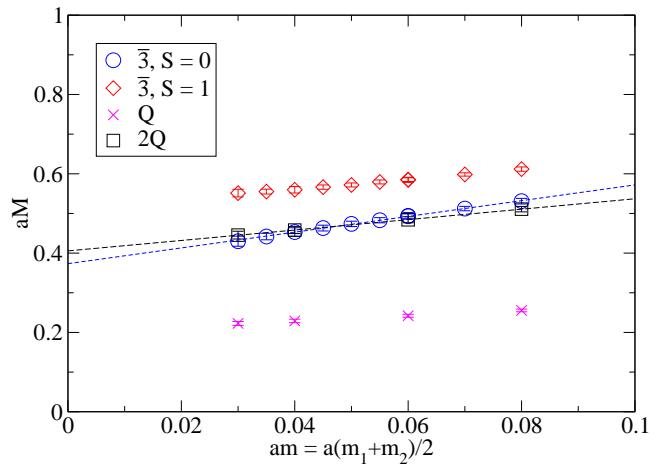}
\caption{\label{fig-dqm}(Color online) Constituent quark and diquark masses
in the Landau gauge, taken from~\cite{Babich:2005ay}.}
\end{figure}

In the above sections, we have observed diquark effects via spatial
correlations in baryon wave functions.  An alternative approach for
investigating diquarks on the lattice is to construct diquark-diquark
correlators and fit their decay in Euclidean time in terms of an effective
``diquark mass''~\cite{Hess:1998sd}.  This is not a gauge-invariant concept
and such a parameter cannot be interpreted as the mass of a physical state,
but it may nevertheless give some indication of the relative strength of
binding.  In~\cite{Babich:2005ay}, we presented results for diquark masses
calculated in the Landau gauge from correlators of the form
\begin{equation}
G(t)=\sum_{\vec{r}} \langle \epsilon_{ijk} u_j(\vec{r},t)
d_k(\vec{r},t) \epsilon_{ij'k'}
\bar u_{j'}(\vec{0},0) \bar d_{k'}(\vec{0},0) \rangle \,,
\label{eq-dqcorr}
\end{equation}
where the indices label color, and implicit spin indices are assigned such
that $u$,$d$ are in either the spin-0 or spin-1 configuration.  In
Fig.~\ref{fig-dqm}, we reproduce a plot taken from~\cite{Babich:2005ay},
showing the dependence of diquark masses on quark mass.  Also included
is the ``constituent quark mass,'' determined by performing a fit to
the quark propagator in the Landau gauge.  If one takes seriously this
``constituent mass'' interpretation, it appears that the scalar diquark may
be bound in the limit of vanishing quark mass.  Here we expand on these
earlier results in two ways, by utilizing non-local sinks in the construction
of the correlators and by reporting values for mass splittings, with errors
taking into account correlations in the data.

In~\cite{Babich:2005ay}, diquark masses were extracted from point-to-point
correlators.  As discussed in Section~\ref{method}, we remain constrained to
point sources, but we are free to use extended sinks, following the approach
applied to mesons in~\cite{Babich:2005ay}.  A natural choice for this purpose
is to use the diquark analogue of the baryon wave functions presented earlier,
again calculated in the Landau gauge.  In particular, we generalize
Eq.~(\ref{eq-dqcorr}) to allow for a separation $r$ between quarks at the sink,
\begin{eqnarray}
G(r,t)&=&\sum_{\vec{r}_u, \vec{r}_d} \langle \epsilon_{ijk} u_j(\vec{r}_u,t)
d_k(\vec{r}_d,t)
\epsilon_{ij'k'} \bar u_{j'}(\vec{0},0) \nonumber \\
&&{}\times\bar d_{k'}(\vec{0},0)
\rangle \delta(r-|\vec{r}_u-\vec{r}_d|) \,.
\end{eqnarray}
We take this function of $r$ at $t=10a$ to define the wave function,
$\varphi(r)=G(r,10a)$, calculating such a $\varphi(r)$ for each state and
quark mass of interest.  Finally, from these we construct an extended-sink
correlator,
\begin{equation}
G_\mathrm{ext}(t)=\sum_r \varphi(r)G(r,t)\,,
\end{equation}
whose fall-off yields the desired diquark mass.

\begin{figure}
\includegraphics*[width=8.5cm]{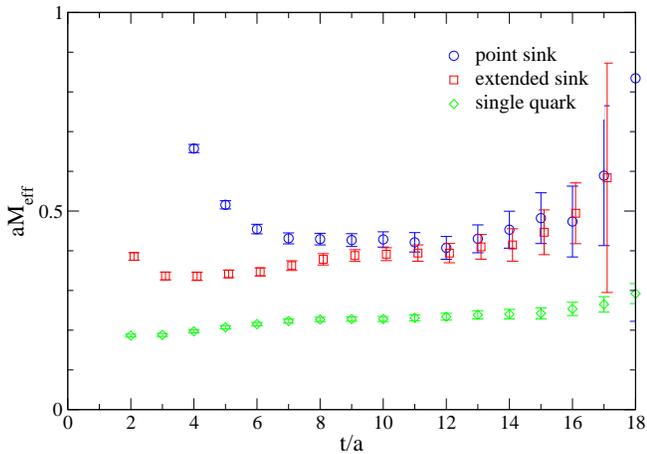}
\caption{\label{fig-s-meff}(Color online) Effective mass of the scalar diquark
as a function of time, calculated with both point and extended sinks.  The
effective quark mass is also shown.}
\end{figure}

\begin{figure}
\includegraphics*[width=8.5cm]{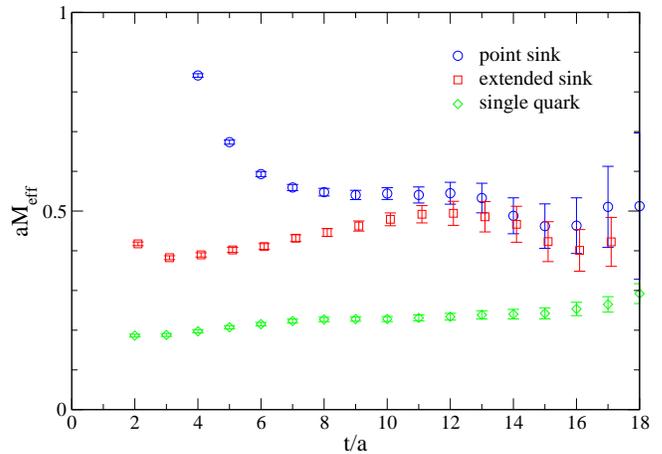}
\caption{\label{fig-v-meff}(Color online) Effective mass of the vector diquark
as a function of time, calculated with both point and extended sinks.  The
effective quark mass is also shown.}
\end{figure}

In Fig.~\ref{fig-s-meff}, we plot the effective mass of the scalar diquark
as a function of time, given by $aM_\mathrm{eff}=-\mathrm{ln}[G(t)/G(t-a)]$,
for both point-sink and extended-sink correlators.  The bare quark mass is
$am_q=0.03$, our lightest value.  The corresponding plot for the vector
diquark is shown in Fig.~\ref{fig-v-meff}.  We find that both point and
extended sink correlators display the same asymptotic effective mass, giving
us confidence that the observed rate of exponential decay may be interpreted
as the ``mass'' of the corresponding diquark in the Landau gauge.  Results
at heavier quark masses exhibit similar behavior.  For the results that
follow, we fit the extended-sink correlators in the region
$11 \leq t/a \leq 14$ and calculate statistical errors by bootstrap.

\begin{figure}
\includegraphics*[width=8.5cm]{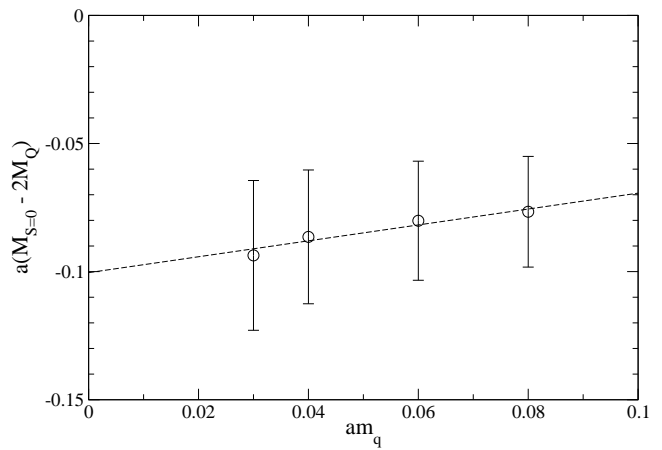}
\caption{\label{fig-dqb}Difference between the mass of the scalar diquark
and twice the constituent quark mass, as a function of bare quark mass.}
\end{figure}

We first consider the ``binding'' of the scalar diquark with respect
to the combined mass of two ``constituent quarks,'' indicated earlier
in~\cite{Babich:2005ay}. In Fig.~\ref{fig-dqb}, we plot the difference
between the scalar diquark mass $M_{S=0}$ and twice the constituent quark
mass $M_Q$.  A naive linear extrapolation gives $a(M_{S=0}-2M_Q)=-0.10(4)$
in the chiral limit.

\begin{figure}
\includegraphics*[width=8.5cm]{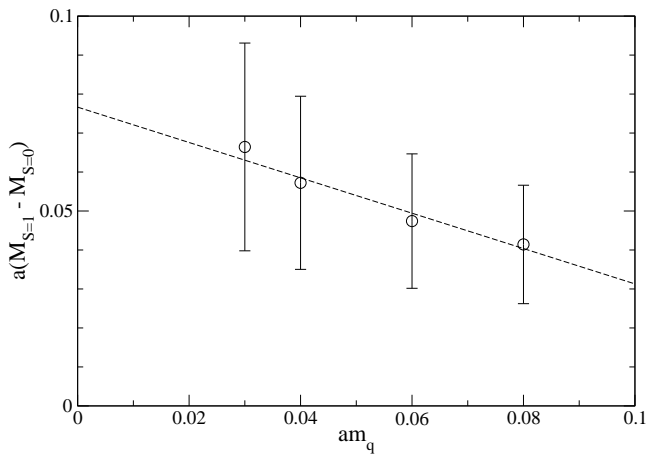}
\caption{\label{fig-dqs}Mass splitting between the scalar and vector diquark,
as a function of bare quark mass.}
\end{figure}

A more robust feature than the binding of the scalar diquark is the
large splitting between it and the vector diquark.  We plot our results for
this mass difference in Fig.~\ref{fig-dqs}.  A linear extrapolation to the
chiral limit gives $a(M_{S=1}-M_{S=0})=0.077(35)$.  Taking $a=2.12$~GeV from
the Sommer scale, we find $M_{S=1}-M_{S=0}=162(75)$~MeV, where the error is
statistical only.  This splitting has also been calculated
in~\cite{Orginos:2005vr,Alexandrou:2006cq} in a gauge-invariant setup where
the free color index of the diquark operator at source and sink is contracted
with a Wilson line.  Equivalently, this scheme corresponds to evaluating the
diquark correlator in a temporal gauge, in which the temporal Wilson line
reduces to the identity.  Our different choice of gauge does not allow a
direct comparison, but we note that the splitting is universally found to be
positive and that it might be interesting to further investigate the gauge
dependence of this quantity.

\section{Conclusions}

In this work, we evaluated baryon wave functions in the Coulomb and Landau
gauges and compared them on the basis of their diquark content.  We found
that spatial correlations were significantly enhanced between quarks in the
scalar diquark configuration as compared to the vector diquark.  Finally,
we presented results for effective mass differences between diquark states
calculated in the Landau gauge.

We acknowledge that our calculation suffers from limitations of the quenched
approximation and the manifest gauge-dependence of our wave functions.
It is encouraging, however, that enhanced correlations were equally pronounced
in both gauges.  It is also noteworthy that in all cases, diquark effects were
found to become more pronounced as quark masses were decreased.
This motivates further, preferably unquenched, calculations at lighter masses.


\begin{acknowledgments}
We thank Robert Jaffe and Frank Wilczek for stimulating discussions
in the early stages of this investigation.  This work is supported in
part by US DOE grants DE-FG02-91ER40676 and DE-AC02-98CH10866, NSF grant
No.~DGE-0221680, EU RTN contracts No.~HPRN-CT-2002-00311 (EURIDICE) and
No.~MRTN-CT-2006-035482 (FLAVIANET), and the CNRS GDR grant n$^\mathrm{o}$~2921
(``Physique subatomique et calculs sur r\'eseau'').  We thank Boston University
and NCSA for use of their supercomputer facilities.
\end{acknowledgments}


\appendix*
\bigskip
\section{Tables of quark separations}

In Tables~\ref{tab-sep-landau} and~\ref{tab-sep-coulomb}, we provide
the root mean square separation between quarks in the various states,
calculated as described in Sec.~\ref{sep}.

\begin{table}[h]
\begin{tabular}{@{\extracolsep{1em}}llllll}
\hline\hline
$am_q$ & \multicolumn{2}{c}{$\Lambda$} & \multicolumn{2}{c}{$\Sigma$} & 
\multicolumn{1}{c}{$\Sigma^*$} \\
\cline{2-6}
& $u-d$ & $u-s$ & $u-d$ & $u-s$ & $u-d$ \\
\hline
0.03 & 5.63(6) & 5.94(5) & 6.03(5) & 5.74(5) & 6.05(5) \\
0.04 & 5.64(5) & 5.93(4) & 6.02(4) & 5.74(4) & 6.04(4) \\
0.06 & 5.63(4) & 5.89(3) & 5.98(3) & 5.72(4) & 6.01(3) \\
0.08 & 5.61(4) & 5.85(3) & 5.93(3) & 5.69(3) & 5.96(3) \\
0.10 & 5.58(3) & 5.81(3) & 5.88(3) & 5.65(3) & 5.92(3) \\
0.25 & 5.30(2) & 5.45(2) & 5.50(2) & 5.35(2) & 5.55(2) \\
0.50 & 4.76(2) & 4.87(2) & 4.91(2) & 4.80(2) & 4.97(2) \\
0.75 & 4.19(2) & 4.30(2) & 4.33(2) & 4.23(2) & 4.39(2) \\
\hline\hline
\end{tabular}
\caption{\label{tab-sep-landau}RMS separation
$\sqrt{\langle |\vec{r}_i-\vec{r}_j|^2\rangle}/a$, in lattice units, between
quarks of flavor $i$ and $j$ as a function of bare quark mass, from baryon
wave functions evaluated at $t=10a$ in the Landau gauge.}
\end{table}

\begin{table}[h]
\begin{tabular}{@{\extracolsep{1em}}llllll}
\hline\hline
$am_q$ & \multicolumn{2}{c}{$\Lambda$} & \multicolumn{2}{c}{$\Sigma$} & 
\multicolumn{1}{c}{$\Sigma^*$} \\
\cline{2-6}
& $u-d$ & $u-s$ & $u-d$ & $u-s$ & $u-d$ \\
\hline
0.03 & 5.17(9) & 5.46(8) & 5.55(8) & 5.27(8) & 5.63(7) \\
0.04 & 5.19(7) & 5.45(6) & 5.53(6) & 5.28(6) & 5.61(6) \\
0.06 & 5.17(5) & 5.41(5) & 5.49(5) & 5.25(5) & 5.56(5) \\
0.08 & 5.14(5) & 5.36(4) & 5.43(4) & 5.22(5) & 5.50(4) \\
0.10 & 5.11(4) & 5.31(4) & 5.37(4) & 5.17(4) & 5.45(4) \\
0.25 & 4.78(3) & 4.92(3) & 4.97(3) & 4.83(3) & 5.04(4) \\
0.50 & 4.26(3) & 4.36(3) & 4.39(3) & 4.29(3) & 4.46(4) \\
0.75 & 3.75(3) & 3.84(3) & 3.87(3) & 3.78(3) & 3.94(3) \\
\hline\hline
\end{tabular}
\caption{\label{tab-sep-coulomb}RMS separation
$\sqrt{\langle |\vec{r}_i-\vec{r}_j|^2\rangle}/a$, in lattice units, between
quarks of flavor $i$ and $j$ as a function of bare quark mass, from baryon
wave functions evaluated at $t=10a$ in the Coulomb gauge.}
\end{table}


\bibliography{bwf}

\end{document}